\documentclass[showpacs,showkeys]{revtex4}

\usepackage{amsmath}
\usepackage{amssymb}
\usepackage{array}
\usepackage{color}

\begin{document}

\title{A nonassociative operator decomposition of strongly interacting quantum fields}

\author{Vladimir Dzhunushaliev
%\footnote{Senior Associate of the Abdus Salam ICTP}
}
\affiliation{Institut f\"ur  Physik, Universit\"at Oldenburg, Postfach 2503, 
D-26111 Oldenburg, Germany, 
\\
Department of Physics and Microelectronic
Engineering, Kyrgyz-Russian Slavic University, Bishkek, Kievskaya Str.
44, 720021, Kyrgyz Republic \\
and \\
Institute of Physics of National Academy of Science
Kyrgyz Republic, 265 a, Chui Street, Bishkek, 720071,  Kyrgyz Republic}
\email[Email: ]{vdzhunus@krsu.edu.kg}

\date{\today}

\begin{abstract}
Working towards an algebra for operators of strongly interacting quantum fields, a nonassociative decomposition of field operators is proposed. In the demonstrated case, quantum corrections appear from the possible bracket permutations. A similarity of these corrections, as compared to corrections from dimensional transmutation, is considered.
%\par\smallskip\noindent 
%{\bf 2000 MSC:} 81R15
\end{abstract}

\keywords{quantum field theory, nonassociative algebra, operator decomposition}

\pacs{11.15.Tk ; 02.90.+p}
\maketitle

\section{Introduction}

One of the most serious problems in modern physics is the quantization of strongly interacting quantum fields. This includes the confinement problem in quantum chromodynamics, quantization of gravitation, and probably also high temperature superconductivity with its strong interaction between Cooper electrons. The problem is that the algebra of quantum operators describing strongly interacting fields is unknown. Known commutation relationships of type
\begin{equation}
\label{1-10}
  \left[ \hat \phi(x), \hat \phi(y) \right] = i \delta(x-y)
\end{equation}
describe free, noninteracting fields (here $ \hat \phi (x) $ is the operator of a free field $ \phi (x) $). From the path integral point of view, we are not able to integrate the action when a nonlinearity exists. In  particular, in quantum chromodynamics we have no analytical methods for integration over a quantum chromodynamical action, in which the interaction potential is of 4$^{th}$ degree of the gauge potentials. For every modern theory aiming at unification of all fundamental interactions, the problem remains.

This article investigates the idea that operators of strongly interacting fields can be decomposed algebraically into products of nonassociative factors. In this case, the nonassociative factors form unobservables, but their product is an observable field operator. From the mathematical point of view, we assume existence of a nonassociative, infinite-dimensional algebra $ \mathbb A $, which contains an associative subalgebra $ \mathbb G \subset \mathbb A $. The observables are modeled to elements of this subalgebra $ \mathbb G $, whereas the quantities modeled on $\mathbb A \setminus \mathbb G $ are not observable in principle. It is also required that the product
$a b \in \mathbb G $, if $a, b \in \mathbb A \setminus \mathbb G $. This article then offers the interpretation that the subalgebra $ \mathbb G $ describes an algebra of strongly interacting fields.                 

\section{Nonassociative decomposition of quantum field operators}

In this section, a nonassociative decomposition of field operators (observables) is constructed, with the goal of identifying an algebra for operators of strongly interacting fields. We assume that the algebra itself will also be nonassociative.

Our construction follows the idea of slave-boson decomposition from condensed matter physics, where it is used to model high-temperature superconductivity (more details on slave-boson decomposition can be found in appendix \ref {slave}). In short, the creation operator of an electron $c ^\dagger _ {i \sigma} $ in a high-temperature superconductor is decomposed into a product of two operators
\begin{equation}
\label{2-10}
  c^\dagger_{i \sigma} = f^\dagger_{i \sigma} b_i,
\end{equation}
where $i$ indicates lattice sites and $\sigma = \uparrow, \downarrow $ is the spin index. The operators in this decomposition, $f ^\dagger _ {i \sigma}, b_i $, are associative.

Following this idea, we now assume that operators of strongly interacting fields $\Phi_m \left (x^\mu \right)$ can be decomposed in similar manner, into nonassociative constituents $f^i_\alpha$ and $b_{i \beta}$:
\begin{equation}
\label{2-20}
  \Phi_m \left( x^\mu \right) = f^i_\alpha \left( x^\mu \right)
  b_{i \beta} \left( x^\mu \right)
\end{equation}
here $m$ is an index where internal and Lorentzian indices are collected, $i$ is the summation index, and the $\alpha$, $\beta$ are contained in $m$ as: $m = \left \{\alpha, \beta \right \}$. Although the constituent operators $f^i_\alpha, b _ {i \beta}$ are not associative, the basic idea in this paper requires their product to model an associative operator. In more mathematical terms, the operators $f^i_\alpha$ and $b _ {i \beta}$ are elements in a nonassociative algebra 
$\mathbb A $, i.e., $f^i_\alpha , b _ {i \beta} \in \mathbb A \setminus \mathbb G$, which contains an associative subalgebra $\mathbb G \subset \mathbb A $, such that $\Phi_m  = f^i_\alpha b_{i \beta} \in \mathbb G$.

It is necessary to note that the operator $\Phi_m$ models observable quantities, whereas the nonassociative $f^i_\alpha, b _ {i \beta}$ are unobservable. For more information on nonassociative operators and unobservables see \cite{Dzhunushaliev:2007cx}.

\section{Examples}

Similar to traditional quantum field theory, where products of field operators are reduced to a normal form, we now consider the product of operators $f^i_\alpha, b _ {i \beta}$ for reduction to an associative normal form. This is achieved exactly when the different nonassociative factors $f^i_\alpha, b _ {i \beta} $ can be expressed through factors similar to Eq. \eqref {2-20}, i.e., as a product of associative operators $\Phi_m $.

Let us consider the following example. We have a product of nonassociative factors:
\begin{equation}
\label{3-10}
  \biggl( 
	  \biggl( 
		  \biggl(
			  f^{i_1}_{\alpha_1} \left( x^\mu_1 \right) 
			  b_{i_1 \beta_1} \left( x^\mu_1 \right) 
			\biggl)
		  f^{i_2}_{\alpha_2} \left( x^\mu_2 \right) 
		\biggl) 
	  b_{i_2 \beta_2} \left( x^\mu_2 \right)
  \biggl)
  \biggl( 
	  \biggl( 
		  \biggl(
			  f^{i_3}_{\alpha_3} \left( x^\mu_3 \right) 
			  b_{i_3 \beta_3} \left( x^\mu_3 \right)
			\biggl)
		  f^{i_4}_{\alpha_4} \left( x^\mu_4 \right) 
		 \biggl) 
  b_{i_4 \beta_4} \left( x^\mu_4 \right)
  \biggl).
\end{equation}
We require to bring this into the form:
\begin{equation}
\label{3-20}
  \biggl(f^{i_1}_{\alpha_1} \left( x^\mu_1 \right) b_{i_1 \beta_1} \left( x^\mu_1 \right) \biggl)
  \biggl(f^{i_2}_{\alpha_2} \left( x^\mu_2 \right) b_{i_2 \beta_2} \left( x^\mu_2 \right) \biggl)
  \biggl(f^{i_3}_{\alpha_3} \left( x^\mu_3 \right) b_{i_3 \beta_3} \left( x^\mu_3 \right) \biggl)
  \biggl(f^{i_4}_{\alpha_4} \left( x^\mu_4 \right) b_{i_4 \beta_4} \left( x^\mu_4 \right) \biggl)
\end{equation}
Since all four terms are associative, this would yield a product of observables:
\begin{equation}
\label{3-30}
  \Phi_{m_1} \left( x^\mu_1 \right) \Phi_{m_2} \left( x^\mu_2 \right)
  \Phi_{m_3} \left( x^\mu_3 \right) \Phi_{m_4} \left( x^\mu_4 \right),
\end{equation}
where 
$\Phi_{m_a} \left( x^\mu_a \right) = f^{i_a}_{\alpha_a} \left( x^\mu_a \right) b_{i_a \beta_a} \left( x^\mu_a \right)$, $a=1,2,3,4$. In order to quantify the difference between Eqs. \eqref{3-10} and \eqref{3-20}, we consider the product
\begin{equation}
\label{3-40}
  \biggl( 
	  \biggl( 
		  \biggl(
			  f^{i_1}_{\alpha_1} \left( x^\mu_1 \right) 
			  b_{i_1 \beta_1} \left( x^\mu_1 \right) 
			\biggl)
		  f^{i_2}_{\alpha_2} \left( x^\mu_2 \right) 
		\biggl) 
	  b_{i_2 \beta_2} \left( x^\mu_2 \right)
  \biggl).
\end{equation}
As $f^{i_a}_{\alpha_a}, b_{i_b \beta_b} \in \mathbb A$ with $ a,b = \left\{ 1,2 \right\} $, a change in bracket arrangement between these nonassociative factors requires additional terms to keep the relation invariant. The corresponding associator is defined as follows:
\begin{equation}
\label{3-50}
  \biggl( 
	  \biggl( 
		  \biggl(
			  f^{i_1}_{\alpha_1} b_{i_1 \beta_1} 
			\biggl)
		  f^{i_2}_{\alpha_2} 
		\biggl) 
	  b_{i_2 \beta_2} 
  \biggl) =
  \biggl(f^{i_1}_{\alpha_1} b_{i_1 \beta_1} \biggl)
  \biggl(f^{i_2}_{\alpha_2} b_{i_2 \beta_2} \biggl) ~ + ~
  \mathrm{Associator}.
\end{equation}
It is interesting to mention that LHS of this equation is obsevable as the RHS involves only 
observables quantities. While we don't know the algebra $\mathbb A$ yet, the associator in general form is then:
\begin{equation}
\label{3-60}
\begin{split}
  \mathrm{Associator} = & 
    \biggl( 
	  \biggl( 
		  \biggl(
			  f^{i_1}_{\alpha_1} \left( x^\mu_1 \right) 
			  b_{i_1 \beta_1} \left( x^\mu_1 \right) 
			\biggl)
		  f^{i_2}_{\alpha_2} \left( x^\mu_2 \right) 
		\biggl) 
	  b_{i_2 \beta_2} \left( x^\mu_2 \right)
    \biggl) -
    \biggl(f^{i_1}_{\alpha_1} \left( x^\mu_1 \right) b_{i_1 \beta_1} 
    \left( x^\mu_2 \right) \biggl)
    \biggl(f^{i_2}_{\alpha_2} \left( x^\mu_2 \right) b_{i_2 \beta_2} 
    \left( x^\mu_2 \right) \biggl) \\
  = &
    - \mu^2_{\alpha_1 \alpha_2 \beta_1 \beta_2 }
\end{split}
\end{equation}
here $\mu^2_{\alpha_1, \alpha_2 \beta_1, \beta_2}$ is a general number in the applicable algebra, and 
$\mu$ can be both real and nonreal.
Therefore,
\begin{equation}
\label{3-100}
  \biggl( 
	  \biggl( 
		  \biggl(
			  f^{i_1}_{\alpha_1} \left( x^\mu_1 \right) 
			  b_{i_1 \beta_1} \left( x^\mu_1 \right) 
			\biggl)
		  f^{i_2}_{\alpha_2} \left( x^\mu_2 \right) 
		\biggl) 
	  b_{i_2 \beta_2} \left( x^\mu_2 \right)
  \biggl)  =
  \Phi_{m_1} \left( x^\mu_1 \right) \Phi_{m_2} \left( x^\mu_2 \right)
  - \mu^2_{\alpha_1 \alpha_2 \beta_1 \beta_2},
\end{equation}
and
\begin{equation}
\label{3-110}
\begin{split}
  &
  \biggl( 
	  \biggl( 
		  \biggl(
			  f^{i_1}_{\alpha_1} \left( x^\mu_1 \right) 
			  b_{i_1 \beta_1} \left( x^\mu_1 \right) 
			\biggl)
		  f^{i_2}_{\alpha_2} \left( x^\mu_2 \right) 
		\biggl) 
	  b_{i_2 \beta_2} \left( x^\mu_2 \right)
  \biggl)  
  \biggl( 
	  \biggl( 
		  \biggl(
			  f^{i_3}_{\alpha_3} \left( x^\mu_3 \right) 
			  b_{i_3 \beta_3} \left( x^\mu_3 \right)
			\biggl)
		  f^{i_4}_{\alpha_4} \left( x^\mu_4 \right) 
		 \biggl) 
  b_{i_4 \beta_4} \left( x^\mu_4 \right)
  \biggl) 
	=
\\
  &
  \Phi_{m_1} \left( x^\mu_1 \right) \Phi_{m_2} \left( x^\mu_2 \right)
  \Phi_{m_3} \left( x^\mu_3 \right) \Phi_{m_4} \left( x^\mu_4 \right) -
  \mu^2_{m_3 m_4} \Phi_{m_1} \left( x^\mu_1 \right) \Phi_{m_2}
  \left( x^\mu_2 \right) -
  \mu^2_{m_1 m_2} \Phi_{m_3} \left( x^\mu_3 \right) \Phi_{m_4}
  \left( x^\mu_4 \right) +
\\
  &
  \mu^2_{m_1 m_2} \mu^2_{m_3 m_4} .
\end{split}
\end{equation}
The indices are combined into $(\alpha_a, \beta_a) = m_a$, in quantities $\mu^2_{\alpha_a \alpha_b \beta_a \beta_b} = \mu^2_{m_a m_b}$. If we apply this reasoning to a nonlinear potential of type $\Phi^4$, we receive:
\begin{equation}
\label{3-120}
\begin{split}
  &
  \biggl(
	  \biggl(
		  \biggl(
		  f^{i_1} \left( x^\mu \right)  b_{i_1} \left( x^\mu \right)
		  \biggl)
		  f^{i_2} \left( x^\mu \right) 
		\biggl)
	  b_{i_2} \left( x^\mu \right)
  \biggl)  
  \biggl( 
	  \biggl(
		  \biggl(
		  f^{i_3} \left( x^\mu \right)  b_{i_3} \left( x^\mu \right)
		  \biggl)
		  f^{i_4} \left( x^\mu \right) b_{i_4} \left( x^\mu \right) 
		\biggl)
  \biggl) =
\\
  &
  \Phi^4 \left( x^\mu \right) - 2 \mu^2 \Phi^2 \left( x^\mu \right) + \mu^4 =
  \left[ \Phi^2 \left( x^\mu \right) - \mu^2 \right]^2.
\end{split}
\end{equation}
The last equation allows to interpret the quantity $\mu$ as a real or imaginary mass, depending on the sign of $\mu^2$. In particular, for $\mu^2 > 0$ it becomes the Mexican hat potential with two global minima. This intriguing result implies that \emph{quantum field theory with strongly nonlinear fields may yield nonperturbative quantum corrections, when field operators are decomposed as the product of nonassociative quantities.}

This result has to be compared to Ref. \cite{coleman}. There, it was shown that radiative corrections could introduce a symmetry breaking ({\it i.e.} negative) mass term into a scalar Lagrangian. This effect is called dimensional transmutation. For the pure scalar case, one cannot rigorously justify a radiatively generated symmetry breaking term, since the scale at which the symmetry breaking occurs lies outside the region where perturbation theory is valid. Nevertheless, it was postulated that a nonperturbative calculation would yield a similar symmetry breaking term, a negative mass, as it happened in our case.

The Coleman-Weinberg mechanism can be applied to the scalar Lagrangian, leading to a spontaneous symmetry breaking potential for $\phi$ of the form:
\begin{equation}
\label{3-130}
	V_{eff} (\phi ) = \frac{\phi ^2}{2 \alpha g^2} + 
	\frac{\phi ^2}{32 \pi ^2}
	\left( \ln \frac{\phi ^2}{{\bar \mu} ^4} - 3 \right)
\end{equation}
here $\bar \mu, g$ are constants. This potential has a nonzero minimum at 
$\phi = \pm v = \pm {\bar \mu}^2 \exp \left(1 - \frac{8 \pi}{\alpha g^2 {\bar \mu}^2} \right)$. In both cases, Eqs. \eqref{3-120} and \eqref{3-130}, we have quantum corrections for the initial potential. 

\section{Acknowledgments} 
I am grateful to the Research Group Linkage Program of the Alexander von Humboldt Foundation for financial support, to J. Kunz for invitation to Universit\"at Oldenburg for research, and to J. K\"oplinger for fruitful discussion. 

\appendix

\section{Some definitions for nonassociative algebras}
\label{na}

For textbook treatment of nonassociative algebras see e.g.~Ref's \cite{schafer}, for applications of nonassociative algebras in physics see e.g.~Ref's.~\cite{okubo1995} and \cite{baez}.

A nonassociative algebra $\mathcal A$ over a field $K$ is a $K$-vector space $\mathcal A$ equipped with a $K$-bilinear map $\mathcal A \times \mathcal A \rightarrow \mathcal A$.

An algebra is unitary if it has a unit or identity element $I$ with $Ix = x = xI$ for all $x$ in the algebra.

An algebra is power associative if $x^n$ is well-defined for all $x$ in the algebra and any positive integer $n$.

An algebra is alternative if $(xx)y = x(xy)$ and $y(xx) = (yx)x$ for all $x$ and $y$.

A Jordan algebra is commutative and satisfies the Jordan property $(xy)(xx) = x(y(xx))$ for all $x$ and $y$.

The associator is defined as follows:
\begin{equation}
  \left( x,y,z \right) \equiv \left( x y \right) z - x \left( y z \right) .
\label{app1-10}
\end{equation}
Any algebra obeying the flexible law
\begin{equation}
  \left( x,y,z \right) = - \left( z,y,x \right)
\label{app1-40}
\end{equation}
is called a flexible algebra.
\par
Any algebra obeying the Jacobi identity
\begin{equation}
 \left[ \left[ x,y \right] , z \right] +
    \left[ \left[ z,x \right] , y \right] +
    \left[ \left[ y,z \right] , x \right] = 0
\label{app1-50}
\end{equation}
is called a Lie-admissible algebra.

\section{Slave-boson decomposition}
\label{slave}

It is widely believed that the low energy physics of High-T$_c$ cuprates (for a review see Ref.\cite{lee}) is described in terms of $t$-$J$ type model, which is given by \cite{LN9221}:
\begin{equation}
	H = \sum \limits_{i,j} J\left(
	{{S}}_{i}\cdot {{S}}_{j}-\frac{1}{4} n_{i} n_{j} \right)
	-\sum_{i,j} t_{ij}
	\left(c_{i\sigma}^\dagger
	c_{j\sigma}+{\rm H.c.}\right),
\label{app2-10}
\end{equation}
where $t_{ij}=t$, $t'$, $t''$ for the nearest, second nearest and 3rd nearest
neighbor pairs, respectively. The effect of the strong Coulomb repulsion is
represented by the fact that the electron operators $c^\dagger_{i\sigma}$ and
$c_{i\sigma}$ are the projected ones, where double occupation is
forbidden. This is written as an inequality
\begin{equation}
	\sum_{\sigma} c^\dagger_{i\sigma} c_{i \sigma} \le 1,
\label{app2-20}
\end{equation}
which is generally difficult to handle. A powerful method to treat this constraint is the so-called "slave-boson" method \cite{B7675,C8435}. In this approach the electron operator is represented as
\begin{equation}
	c^\dagger_{i\sigma} = f_{i\sigma}^\dagger b_{i}
\label{app2-30}
\end{equation}
where $f_{i\sigma}^\dagger$, $f_{i \sigma}$ are the fermion operators, while $b_{i}$ is the
slave-boson operator. This representation, together with the constraint
\begin{equation}
	f_{i\uparrow}^\dagger f_{i\uparrow} + f_{i\downarrow}^\dagger f_{i\downarrow} +
	b^\dagger_{i} b_{i} = 1,
\label{app2-40}
\end{equation}
reproduces the entire algebra of electron operators. The physical meaning of operators $f$ and $b$, however, is unclear: Do these fields exist in nature or not?

\section{Spin-charge separation}
\label{spin}

It is proposed in Ref. \cite{ref:splitting:niemi} to split gluons in the same manner as slave-boson decomposition from high-$T_c$ superconductivity models. This splitting is based on the field decomposition~\cite{ref:splitting:faddeev}, which is applied to the off-diagonal gluons while leaving the diagonal
gluons intact. In SU(2) Yang--Mills theory, the splitting
of the off-diagonal gluons $A^{1,2}_\mu$ \cite{ref:splitting:niemi,ref:splitting:faddeev},
\begin{eqnarray}
	A^1_\mu + i A^2_\mu &=& \psi_1 {\vec e}_\mu + \psi^*_2 {\vec e}^*_\mu,
\nonumber\\
	{\vec e}_\mu {\vec e}_\mu &=& 0,
\nonumber\\
	{\vec e}_\mu {\vec e}^*_\mu &=& 1,
\label{app3-10}
\end{eqnarray}
leads to appearance of two electrically charged (with respect to the Cartan subgroup
of the color gauge group) Abelian scalar fields $\psi_{1,2}$, and the electrically neutral field ${\vec e}_\mu$, which is a complex vector.

In Ref. \cite{chernodub}, a generalization of the spin-charge decomposition
in high-$T_c$ superconductors \eqref{app2-30} to SU(2) Yang--Mills theory is proposed. This decomposition splits the SU(2) gluon field into spin and color degrees of freedom, treating
all color components equally:
\begin{equation}
	A^a_\mu(x) = \Phi^{ai}(x) \, e^i_\mu(x).
\label{app3-20}
\end{equation}
Here, $\Phi^{ai}(x)$ is the $3\times 3$ matrix, and $e^i_\mu(x)$ are the three vectors forming an (incomplete) orthonormal basis within the four dimensional space-time, $e^i_\mu(x) e^j_\mu(x) = \delta^{ij}$. The elements of $\Phi^{ai}(x)$ and $e^i_\mu(x)$ are real functions that are labeled by color ($a=1,2,3$), as well as internal ($i=1,2,3$) and Euclidean vector ($\mu=1,\dots,4$) indices. Obviously, Eq.~\eqref{app3-20} is a color-symmetric generalization of Eq.~\eqref{app3-10}. A similar decomposition for the SU(3) gauge theory is made in Ref. \cite{Dzhunushaliev:2006sh}.


\begin{thebibliography}{99}

\bibitem{Dzhunushaliev:2007cx}
V.~Dzhunushaliev,
%``Observables and unobservables in a non-associative quantum theory,''
J. of Generalized Lie Theory and Applications, \textbf{Vol. 2}, No. 4, 269-272 (2008);
hep-ph/0702249.

\bibitem{coleman}
S. Coleman and E. Weinberg,
%``Radiative corrections as the origin of spontaneous symmetry breaking'',
Phys. Rev. \textbf{D7}, 1888 (1973).

\bibitem{schafer}
R. Schafer.
Introduction to Non-Associative Algebras.
Dover, New York, 1995;\\
T. A. Springer and F. D. Veldkamp.
Octonions, Jordan Algebras and Exceptional Groups,
Springer Monographs in Mathematics, Springer, Berlin, 2000.

\bibitem{okubo1995}
Susumu Okubo.
Introduction to Octonion and Other Non-Associative Algebras in Physics.
Cambridge University Press, Cambridge, 1995.

\bibitem{baez}
J.~C.~Baez.
The Octonions.
Bull. Amer. Math. Soc., \textbf{39} (2002) 145-205;
math.ra/0105155.\\
J. L$\tilde{o}$hmus, E. Paal and L. Sorgsepp.
About Nonassociativity in Mathematics and Physics.
Acta Appl. Math., \textbf{50} (1998) 3-31.

\bibitem{lee}
P.A. Lee, N. Nagaosa and X.-G. Wen,
``Doping a Mott insulator: physics of High Temperature superconductivity'',
Rev. Mod. Phys. \textbf{78}, 17 (2006);
cond-mat/0410445.

\bibitem{LN9221}
P.~A. Lee, N. Nagaosa,
``Ginzburg-Landau theory of the spin-charge-separated system'',
Phys. Rev. B, \textbf{45}, 966 (1992).

\bibitem{B7675}
S.~E. Barnes,
``New method for the Anderson model'',
J. Phys. F, \textbf{6}, 1375 (1976).

\bibitem{C8435}
P. Coleman, Phys. Rev. B,
``New approach to the mixed-valence problem'',
\textbf{29}, 3035 (1984).

\bibitem{ref:splitting:niemi}
A.~J.~Niemi,
%``Dual superconductors and SU(2) Yang-Mills,''
JHEP {\bf 0408}, 035 (2004); \\
A.~J.~Niemi and N.~R.~Walet,
%``Splitting the gluon?,''
e-print hep-ph/0504034.

\bibitem{ref:splitting:faddeev}
L.~D.~Faddeev and A.~J.~Niemi,
%``Electric-magnetic duality in infrared SU(2) Yang-Mills theory,''
Phys.\ Lett.\ B {\bf 525}, 195 (2002).

\bibitem{chernodub}
M.~N.~Chernodub,
%``Yang-Mills theory in Landau gauge as a liquid crystal,''
Phys.\ Lett.\ B {\bf 637} (2006) 128.

\bibitem{Dzhunushaliev:2006sh}
V.~Dzhunushaliev,
``Spin-charge separation for the SU(3) gauge theory,''
Int.\ J.\ Mod.\ Phys.\  A {\bf 21}, 6457 (2006); 
hep-ph/0607312.


\end{thebibliography}
\end{document}